# Compact Metasurface Terahertz Spectrometer

**Wenye Ji[1, #], Jin Chang[2, #, *], Behnam Mirzaei[1], Marcel Ridder[3], Willem Jellema[3], Wilt Kao[4], Alan Lee[4], Jian Rong Gao[1,3], Paul Urbach[1], Aurèle J.L. Adam[1].**

1.  Department of Imaging Physics, Delft University of Technology, Lorentzweg 1, 2628 CJ, Delft, The Netherlands

2.  Department of Quantum Nanoscience, Delft University of Technology, Lorentzweg 1, 2628 CJ, Delft, The Netherlands

3.  SRON Netherlands Institute for Space Research, Niels Bohrweg 4, 2333 CA Leiden and Landleven 12

9747 AD Groningen, The Netherlands

4. LongWave Photonics LLC, Mountain View CA 94043, USA

#These authors contribute equally to the work
*Corresponding author: j.chang-1@tudelft.nl

## Abstract:

The electromagnetic spectrum in the terahertz frequency region is of significant importance for understanding the formation and evolution of galaxies and stars throughout the history of the universe and the process of planet formation. Within the star forming clouds the constituent atoms and molecules are excited to produce characteristic emission and absorption lines, many of which happen at the terahertz frequencies. Thus, detecting the spectral signatures as unique fingerprints of molecules and atoms require terahertz spectrometers, which need to be operated in a space observatory because of the water vapor absorption in the earth atmosphere. However, current terahertz spectrometers face several challenges that limit their performances and applications, including a low resolution, limited bandwidth, large volume, and complexity. In this paper, we address the last two issues by demonstrating a concept of a compact terahertz spectrometer using metasurface. We start by modelling, designing, and fabricating a metasurface, aiming to optimize its performance within a band from 1.7 to 2.5 THz. Next, we make use of an array of quantum cascade lasers that operate at slightly different frequencies around 2.1 THz to validate the performance of the spectrometer. Finally, we apply the spectrum inversion method to analyse the measured data to confirm a resolution *R* (*f*/Δ*f*) of at least 273. Our results demonstrated a miniaturized terahertz spectrometer concept successfully. Our findings and provide a novel solution for the future spectrometer design, which can also be applied in other optical wavelengths.

## 1. Introduction:

Spectroscopical observations play an important role for space astronomy in the terahertz (THz) frequencies or at Far-Infrared (FIR) wavelengths, where the radiation could provide answers to some key questions on the formation and evolution of galaxies and stars throughout the history of the universe and the process of planet formation. It is known that the temperatures and densities within the star forming clouds excite the constituent atoms and molecules to produce emission and absorption lines, many of which happen at THz frequencies. Furthermore, the long wavelength nature of THz radiation allows for a study of the active interiors of galaxies throughout cosmic history because galaxies are inherently dusty, where dust absorbs the short, visible wavelength light and re-radiates the energy in the far-IR [1].

A spectrometer forms the heart of many astronomic instruments at THz [1]. Typically, one applies grating spectrometers or Fourier transformation spectrometers (FTS) spectrometers as low and medium resolution spectrometers, with a spectral resolution (R) of $10^2$-$10^4$, while heterodyne receivers as high-resolution spectrometers can reach an R of higher than $10^6$. The R is defined as the average wavelength divided by the wavelength interval. The choice of a spectrometer depends on the type of observations and science goals [2]. The Herschel Space Observatory, launched in 2009 by ESA, had all the three types of spectrometers on board, covering the wavelength span between 60 and 700 µm [3]. The observatory must be operated from space because the Earth's atmosphere is opaque over most of the FIR wavelengths.

Grating spectrometers are in particular interesting because of their sensitivity, resolution, bandwidth and imaging capability when they are combined with a large, low noise detector array of more than $10^3$ pixels. They are highly demanded to study distant galaxies with extremely faint signals, where R of $10^2$-$10^4$ is sufficient [4]-[6]. A few space mission concepts, such as SAFARI-SPICA [4], the Origins Space Telescope (OST) [5], and Galaxy Evolution Probe (GEP) [6], have proposed to use grating spectrometers. In contrast to Fourier Transform Spectrometers (FTS), grating spectrometers acquire spectral data by diffracting light through a grating without the involvement of movable components, rendering them mechanically uncomplicated. A typical THz grating spectrometer consists of a slit, collimator, reflective mirror, reflective grating, concave mirror, and detectors [6]-[7].

However, the physical size and mass of optical components of a grating spectrometer, where the resolution is determined by the optical path difference, are still an issue for space applications. For example, four grating modules for SAFARI-SPICA with an R of 300 to cover a wavelength range between 34 and 230 µm, the maximal physical size of a module varies from 0.4 to 0.5 meters [4], [6]. The spectrometer becomes even more bulky and heavier or has more complexities by adding a Fourier transform module when it aims to have a higher resolution or to operate at longer wavelengths [4]. In order to meet the requirements of forthcoming space observatories operating in the far-infrared (FIR) wavelengths, the development of a compact terahertz (THz) spectrometer with both high resolution and a wide bandwidth is imperative.

In the past decade, two dimensional (2D) planar metamaterials, known as metasurfaces, have attracted enormous amount of research interests [8]-[21]. Plentiful captivating metasurface devices spring up, including vortex generators [22], [23], achromatic devices [24]-[26], multi-functional devices [27]-[29]. As a metasurface can encompass both dispersion and focusing functionalities within a single element, it holds promise as a viable replacement for conventional, bulky components such as gratings and reflective mirrors.

Several works on miniaturization of spectrometer structures by introducing a metasurface has been reported in literature. However, most of them were demonstrated at visible or other short wavelengths [30]. For example, M. Faraji-Dana et al [31] proposed a concept of a compact spectrometer by folded metasurface optics made from a 1-mm-thick glass slab with a volume of 7 mm$^3$. The resolution R equals 680 for the wavelengths from 760 nm to 860 nm. Nevertheless, the efficiency is 25% and is thus very low because of the multi-reflection between metasurfaces and dielectric loss. Thus, this architecture is not applicable to THz wavelengths. A. Y. Zhu et al. [32] proposed a compact aberration-corrected spectrometer by a dispersive metasurface. The resolution R is 600 within a wavelength span between 450 and 700 nm. However, the efficiency was generally low because the focus point of the lens is in off-axis. An alternative approach by A. Endo et. al. [33] is to realize a spectrometer on a chip, namely wide band radiation is spectrally separated by introducing superconducting filter banks. It was demonstrated at a much longer wavelength of 860 μm. Such a miniaturized spectrometer is limited to the working wavelengths that are roughly longer than 300 μm because of the use of available lossless superconductors. Until now, no practical compact spectrometers and in particular no metasurface based spectrometers have been reported at THz or at FIR wavelengths.

In this paper, we demonstrate a novel type of THz spectrometer, the core of which is a single metasurface. The schematic of our concept is illustrated in Figure 1. The THz wave in space is incident into the spectrometer (for simplicity, no telescope was included) . For the radiations of different wavelengths, they are diffracted and focused on different position in the focusing plane. We begin by modelling, then design and fabricate a metasurface sample, with the goal of optimizing its performance within a frequency span of 1.7 to 2.5 THz. Next, we use an array of quantum cascade lasers (QCLs) that operate at slightly different frequencies around 2.1 THz as input sources to validate our device performance by measuring resolution and efficiency of the spectrometer. Finally, we employ the spectral inversion algorithm to analysis experimental data and achieve a resolution R of 273. Our results have proved for the first time a compact THz metasurface spectrometer. This advance opens a new avenue towards astronomic instruments for future space observatories.

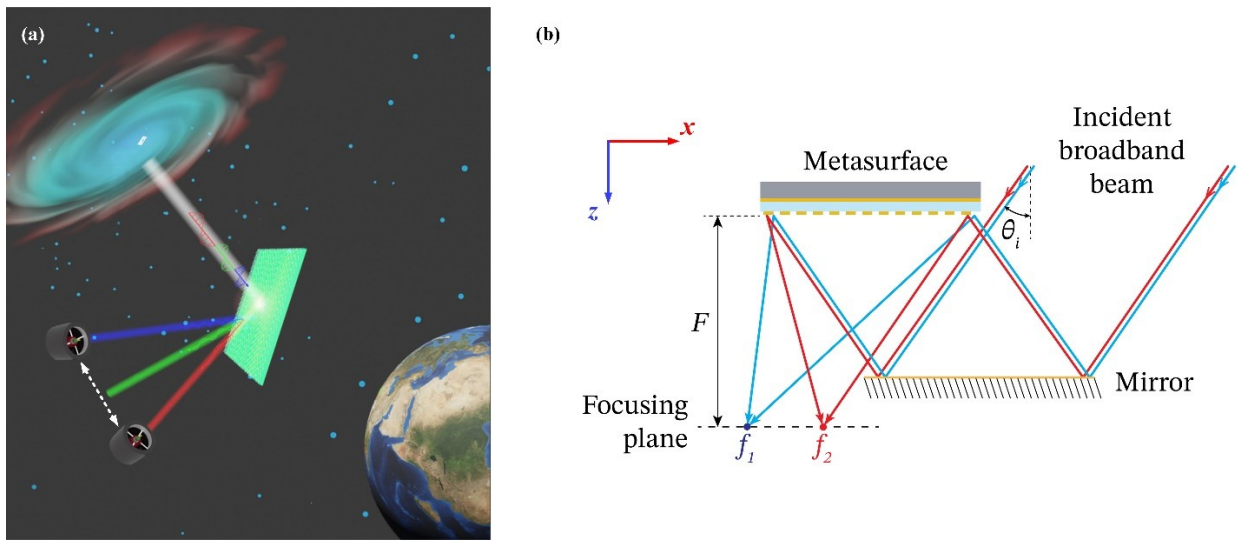

Figure 1. Schematic of a compact metasurface terahertz spectrometer. (a) The THz wave in space is incident into the spectrometer. For the lights of different wavelengths, they are diffracted and focused on different positions in the focusing plane. (b) The detailed schematic of our terahertz metasurface spectrometer concept.

## 2. Metasurface spectrometer concept

We propose the concept of a THz spectrometer system in Figure 1(b). The system contains one flat, reflection mirror and one metasurface that possessing both diffraction and focusing functions. The incident light is a broadband, plane wave with an oblique angle $\theta_i$ of 30°. Incident light is reflected by the flat mirror onto the metasurface. Depending on the wavelengths, light is diffracted and focused to different positions in the focusing plane. Here, we designed the metasurface working in reflection mode rather than transmission because in transmission, the efficiency is much lower than the reflection and working bandwidth is also narrower [34], [35]. The free and front views of a metasurface unit cell are shown in Figure 2(a) and (b), respectively. The unit cell consists of three layers on a substrate: the bottom layer is 300-nm-thick gold (yellow colored) with a conductivity of $3.3 \times 10^7$ S m$^{-1}$, which leads to a skin depth of less than 60 nm at 2 THz and ensures no transmission, but a high diffraction efficiency for the THz waves. The middle layer is a dielectric made of polyimide (blue colored) with a relative permittivity of 3.3 and loss tangent of $4 \times 10^{-3}$ [36]. The top layer (on the polyamide) is also gold with a thickness of 100 nm and forms periodic double-anchor like structures (yellow colored).

To realize broadband and highly efficient diffraction, we select the following initial values for the parameters for a double-anchor as the unit cell. The dimension $p$ of the unit cell is 45 μm, which also defines the periodicity of the metasurface. The parameters of the 'shank' include a radius ($r$) of 21 μm, a width ($w$) of 3 μm, an angle ($\alpha$) defining the length of the arm at 75°, and an angle ($\beta$) defining the orientation at 45°. The thickness $d$ of the polyimide is 19 μm. The electromagnetic behavior of the double-anchor is characterized using CST Microwave Studio. In the simulation, the Frequency Domain Solver is used, and periodic boundary conditions are applied around the unit cell. The incident wave propagates with the oblique angle 30° relative to -z-axis. We use Reflection Jones Matrix [37] $J = \begin{bmatrix} r_{xx} & r_{xy} \\ r_{yx} & r_{yy} \end{bmatrix}$ to describe reflection waves, where $J$ is the reflection matrix, and $r_{ij}$ represents the reflection coefficient of $i$-polarized when incidence is $j$-polarized wave with $i$ and $j$ being either $x$ or $y$. We consider only the reflection here. The reflection properties of the double-anchor unit cell are shown in Figure 2(c). If the unit cell is lossless, both $|r_{yx}|^2 + |r_{xx}|^2$

= 1 and $|r_{xy}|^2 + |r_{yy}|^2 = 1$ are held. We mainly concentrate on cross-polarization $r_{xy}$, the coefficient of which is found to be above 0.8 across the whole frequency span. The phase property of the structure is determined by $\beta$ and $\alpha$. We give an example of the simulation results based on the unit cell working at 2.1 THz. In the left part of Figure 2(d), if $\beta$ is -45° and $\alpha$ changes from 20° to 70°, it can realize the reflection phase coverage from $2\pi$ to $\pi$. In the right part, if $\beta$ is 45° and $\alpha$ is varied, the structure can realize the phase coverage from $\pi$ to 0. In this way, the full $2\pi$ phase coverage is possible, which can be used to design the final metasurface. Furthermore, the reflection efficiency of such a unit cell is always above 90 % at 2.1 THz for different values of $\alpha$ and $\beta$ [38], [39].

To study the bandwidth of the diffraction response, we plot different phase and amplitude response curves in figure 2(e), where $\beta$ switching from -45° to 45° with $\alpha$ varying as well. The structure can work well between 1.7 to 2.5 THz. For all the phase curves, they are parallel to each other, implying that within the entire band, the structures have a constant dispersion relationship [24]. In addition, within the band, all the reflection coefficients are always above 0.85, that ensures a high efficiency of the metasurface.

Now, we introduce a spectrometer configuration based on the metasurface as shown in Figure 2(f). Firstly, the metasurface in $x$ direction consists of 250 double-anchor unit cells with variable, but required $\alpha$ and $\beta$, which cover a length of 11.25 mm. In the $y$-direction, the same unit cells are repeated 250 times, leading to the same size as in the $x$-direction. Because of such a configuration, here we define it as a 1D meta-surface structure. The focal length $F$ of the metasurface is set to be 15 mm. The distance between the metasurface and the flat mirror is 13 mm. We set the zero of the $x$-coordination in the left edge of the metasurface and $y$-coordination in the bottom edge of the metasurface. In theory we target the focused spot at position of 5.5 mm in $x$-direction in focusing plane for 2.1 THz. The phase design of the unit cells in the metasurface is according to the constructive interference principle. As shown in Figure 2(f), the optical path distance between trace A and B should be the same. Then, resulted phase response, which is the phase shift between the incident wave (to a unit cell) and deflective wave from the unit cell, is determined for every unit cell by the following formula [8],

$$\phi_i(f) = \phi_1(f) + \frac{2\pi f}{c}\left(\sqrt{[x_i - x_0(f)]^2 + F^2} - \sqrt{[x_1 - x_0(f)]^2 + F^2}\right) + \phi_{delay}, \tag{1}$$

where, as marked in Figure 2(f), $\phi_i(f)$ is the phase required for unit cell $i$ at frequency $f$, $\phi_1(f)$ the phase of the first unit cell in the left edge of the metasurface, $x_0(f)$ is the position of the focused beam in the focusing plane. $c$ is the light speed and $x_i$ is the centre position for the unit cell $i$. $\phi_{delay}$ is shown in the figure schematically. Because of the plane wave assumption, the equal phase plane is perpendicular to the propagation direction of the incident wave, thus, $\phi_{delay} = x_i \sin\theta_i \frac{2\pi f}{c}$. Finally, to satisfy the required phase in Eq. (1), we can select a double-anchor by choosing proper $\alpha$ and $\beta$, to fill in a unit cell in the metasurface. The variable parameters $\alpha$ and $\beta$ for every unit cell can be found in **Supporting Information Section 1**.

Next, we perform simulations for the spectrometer configuration using CST Microwave studio, where the Time-Domain Solver is again applied. The incident light is broadband plane wave, with the electric field in *y*-direction, and covers the full area of the metasurface. We calculate the intensity distribution of *x*-polarized field for a few different wavelengths in the focusing plane. The results are plotted in Figure 2(g). The peak position of diffracted beam for 2.10 THz, 2.11 THz, 2.20 THz, and 2.30 THz is at 5552 μm, 5535 μm, 5384 μm, and 5227 μm in *x*-axis, respectively. Simulated half power beam width for 2.10 THz, 2.11 THz, 2.20 THz, and 2.30 THz is 162 μm, 164 μm, 177 μm, 190 μm, respectively. Furthermore, peak positions within 0.01 THz at 2.1 THz are distinguishable, thus the spectral resolution R of at least 210. We also calculate the intensity distribution of the co-polarization reflection wave in the focusing plane (by assuming a

*y*-polarized incident wave) from the metasurface, which is discussed in **Supporting Information Section 2.**

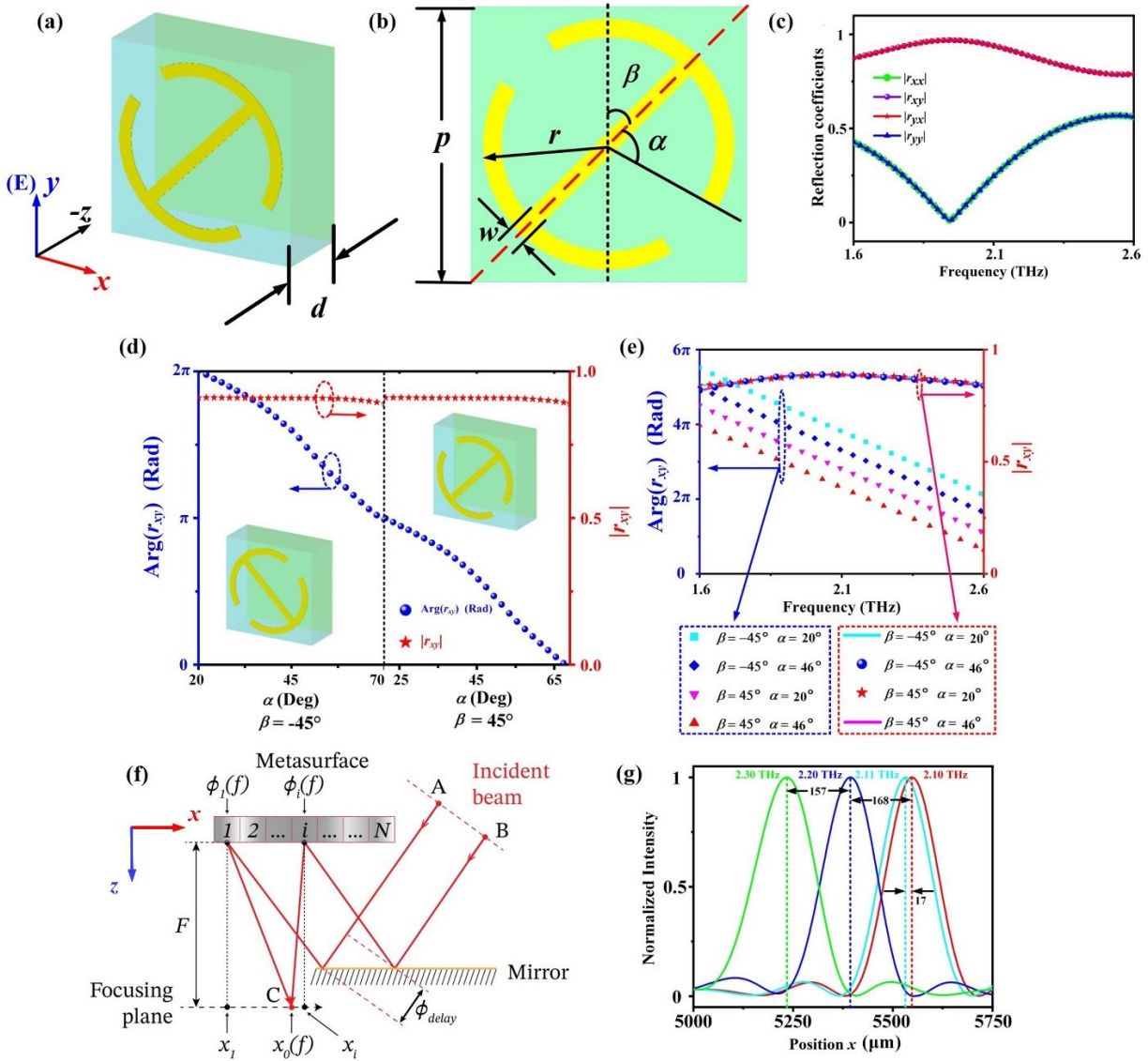

Figure 2. Metasurface spectrometer concept, design and simulation. (a) Free view of a double-anchor unit cell for a metasurface. (b) the front view of the double-anchor unit cell. (c) The reflection properties of a unit cell for initial parameters as a function of the frequency. (d) The phase and amplitude properties for different unit cell parameters at 2.1 THz. (e) Broadband phase and amplitude response curves for parameters $\beta$ switching from -45° to 45° $\alpha$ switching from 20° to 46°. (f) Schematic of final design for metasurface spectrometer configuration. (g) Normalized intensity distribution for different wavelengths in the focusing plane.

## 3. Fabrication and experimental set-up

The fabrication of the metasurface consists of three key steps. Firstly, an Au layer of 300 nm acting as a mirror is deposited on a silicon wafer. Then, a polyimide layer is spin-coated on top of the Au with 19 μm thickness. Second Au layer of 100 nm is deposited atop the polyimide and is subsequently patterned by photolithography and etching to form periodic double-anchor structures. The schematic of the cross section of a completed metasurface sample and an optical micrograph of its surface are shown in Figure 3(a).

We have built an experimental setup shown in Figure 3(b) to characterize the spectrometer, where the metasurface structure on a holder, together with the flat mirror, is shown in Figure 3(c). Before we explain how the experiment was done, we first introduce THz Quantum Cascaded Lasers (QCLs) as calibration sources. The distributed feedback (DFB) laser array (SN556) provided by LongWave Photonics contains 8 individual lasers on one chip (see Figure 3(c)) with nominal peak power of 1 mW at a working temperature of 49 K (see Figure 3(e)), emitting the radiation in the range of 2.11 to 2.19 THz as shown in Figure 3(d). The gain medium used for the QCLs is based on Resonant-Phonon (RP) depopulation [40] scheme with a two-level injector. The frequency selectivity of the lasers is provided by DFB structure based on the antenna-coupled wire laser design as demonstrated in [41], [42]. All lasers in the array maintain single-mode operation. The advantages associated with Quantum Cascade Lasers (QCLs) encompass features such as coherent emission characterized by an exceptionally narrow linewidth, typically less than 1 MHz [41], as well as a robust output power profile. The high output power capability of QCLs facilitates the utilization of room temperature detectors, thereby significantly streamlining the experimental setup.

To obtain the position of a diffracted/focused beam in the focusing plane, we use a pyro-electric detector (in Fig. 3(b)) with a NEP (Noise-equivalent power) of $5\times10^{-9}$ WHz$^{-1/2}$ and with a pinhole of 100 μm in diameter. The detector, mounted on a computer controlled 2D translation stage, is used to map the intensity of diffracted beam by raster scanning in a step of 100 μm along the focusing plane. However, when we do only 1D fine scanning, we reduce the pinhole to 30 μm, while scan in a reduced step of 10 μm. We make use of only three QCLs in our experiment: D4 at 2.150 THz, D7 at 2.180 THz, and D8 at 2.188 THz, where the frequencies were determined separately by an FTS. The frequency difference between D7 and D8 is in particular interest because it is only 8 GHz, which is the smallest in the chip. We target to distinguish them (D7 and D8 lines) to demonstrate a high spectral resolution. Furthermore, these three lasers have higher output power than the rest. We operate the QCLs in a Stirling cooler. More information over the QCLs is given in **Supporting Information Section 3 Figure S2**.

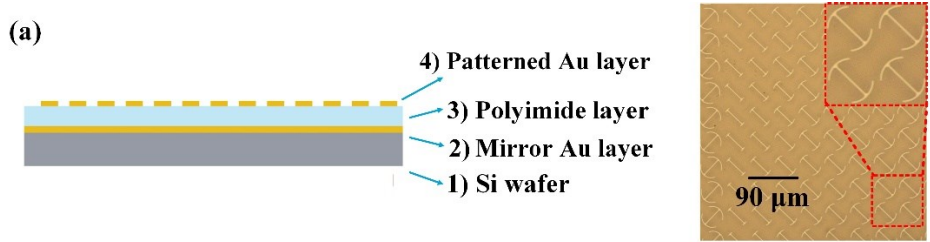

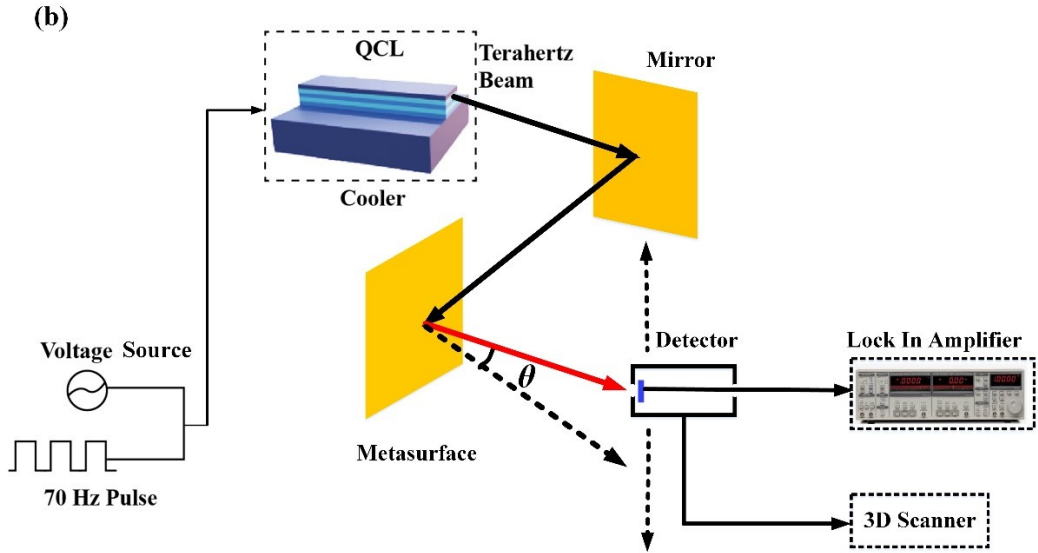

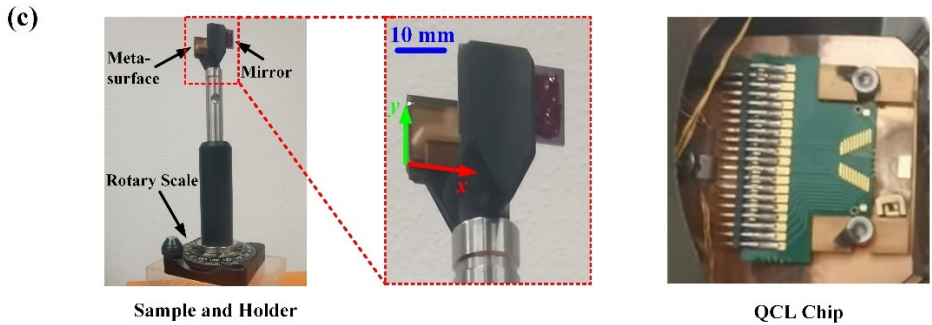

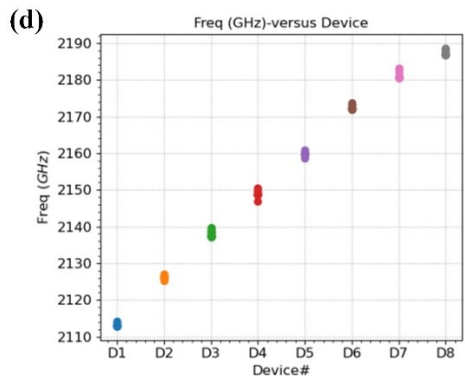
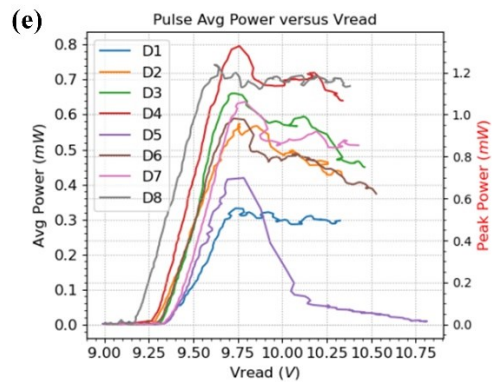

Figure 3. Metasurface structure and experimental set-up. (a) Schematic cross-section of the metasurface structure and fabricated sample. (b) Schematic overview of the entire experimental setup for characterizing a metasurface spectrometer system. (c) Pictures of a metasurface sample and a flat mirror on a holder and the chip containing the QCLs. (d) The frequencies for the 8 QCLs. (e) Average power curves against voltage source for 8 different QCLs on the chip.

## 4. Experimental results and data analysis with spectrum inversion method

QCL beams are not ideal, deviated considerably from a Gaussian beam, that is partly due to the QCL itself and partly due to non-optimized optics inside the cooler. To determine the peak position of the diffracted beam accurately, we use a diaphragm with a variable aperture from 4 mm to 2.5 mm in diameter to filter the beam to a circular, symmetric beam. The method is described in **Supporting Information Section 4**. We also note that a change of incident beam size does not influence the focus and deflection effect of the metasurface, but can affect the focused spot size because of the change of the numerical aperture (NA). The simulated intensity distribution in focusing plane for different beam size of the incidence at 2.10 THz is discussed in **Supporting Information Section 5.**

We begin with a measurement of the metasurface at 2.150 THz using QCL D4. We first characterize the THz beam using a mirror instead of the metasurface, with the same dimension and at the same position. The schematic of measurement configuration is illustrated in Figure 4(a). The beam, measured at the focusing plane, is shown in Figure 4(b). We find the measured beam in this case is 4 mm in diameter. Because of only the mirror, the measured beam and position reveal the properties of the input beam, while the beam position is purely due to the dispersionless reflection. Then, we replace the mirror with the metasurface as shown in Figure 4(c) and repeat the measurement. The measured beam at the focusing plane, which is shown in Figure 4(d), shows that the diffracted beam shifts to the right with a distance of 7.6 mm with respect to the reflected beam and the beam becomes much narrower in the *x*-direction. In the same plot the zero-order diffraction is visible, but has a much lower intensity. Apparently, we have observed a signature of the 1D metasurface, which concurrently diffracts and focuses the beam at only horizonal direction. To quantify the diffracted beam and compare it with the simulations, we make 1D measurements of the two cases by fine scanning the detector in the *x*-direction along the centre of the beam, as indicated by the yellow dot line. The results are summarized in Figure 4(e). The beam (centre) shifts from -3.31 mm to 5.09 mm. The half power beam widths of reflected beam and diffracted beam (in the *x*-direction) are 4.03 mm and 0.37 mm, respectively. The relative shift in distance will be compared to the simulation later. We now compare measured diffracted beam profile with our modelling in Figure 4(f), where we found that the measured beam has a similar profile as the one from the simulation, except the width is 70% wider. This difference is mainly due to the narrower incident beam (of 4 mm) in the experiment than full illumination to the metasurface in the simulation, which leads to a smaller NA and thus not perfectly focused on the focusing plane. We also estimate the efficiency of the metasurface sample, which is defined as the ratio between the power of incoming beam after the aperture and that of diffracted one. The relative power of the incident beam after the aperture is determined by integrating the total intensity within the beam area, while the power of the diffracted beam is estimated by integrating the intensity of the beam in Figure 4(d). The efficiency, thus the ratio between the two, is 78.4%. The loss comes mainly from the zero-order diffraction, which is known to be 12.7%, and then resonant loss from double-anchor gold structures, the absorption loss from the polyimide [36], and wave scattering in free space add up to 8.9 % loss.

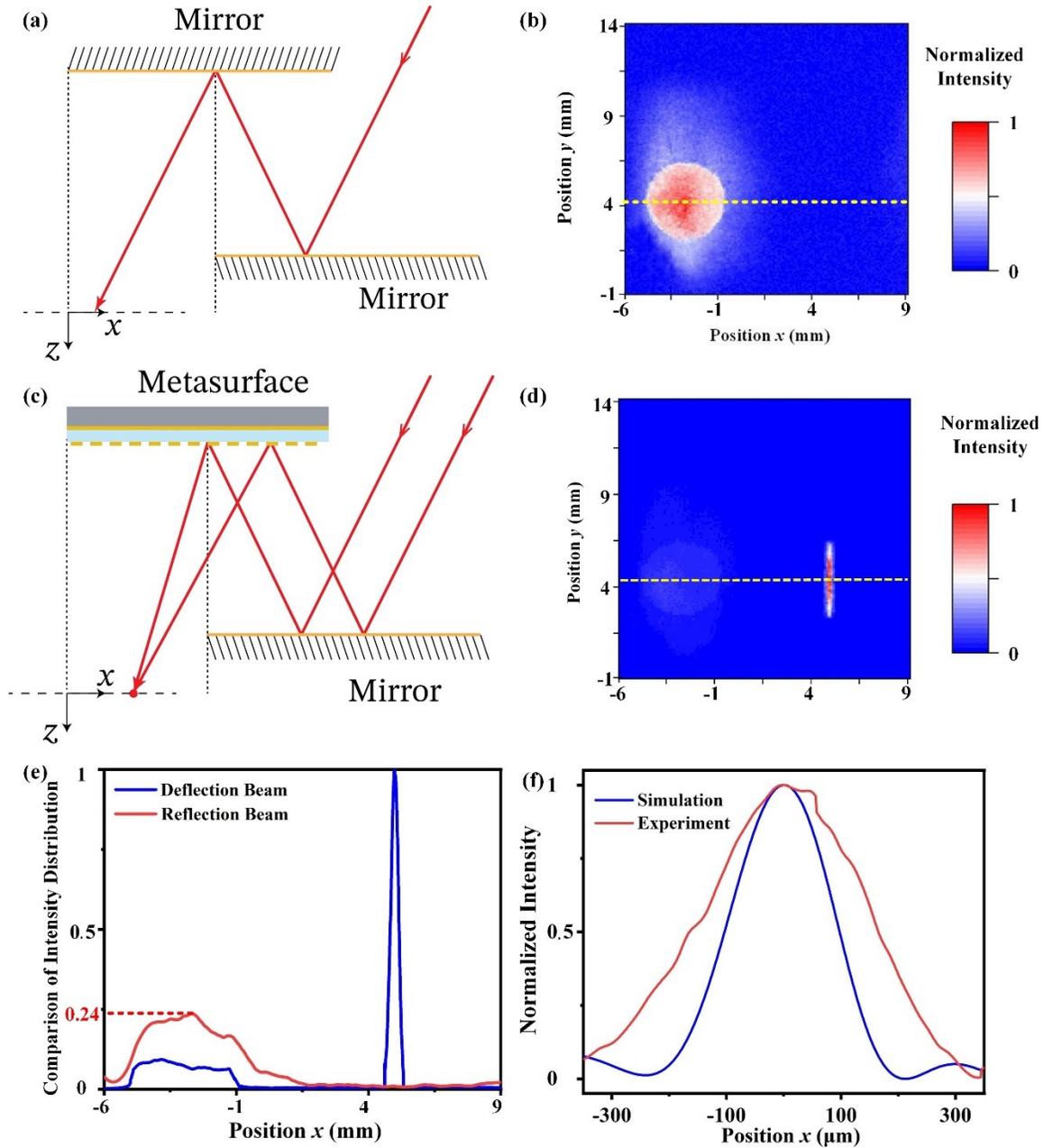

Figure 4. Measurement results of the metasurface spectrometer structure at 2.15 THz. (a) Schematic of the setup when the metasurface is replaced by a mirror. (b) Measured reflection beam using the setup in (a). (c) Schematic of the metasurface spectrometer set up. (d) Measured deflection beam using the setup in (c). (e) Detailed intensity distributions of the reflection- and deflection-beam in 1D scanned along the dashed lines indicated in (b) and (d). (f) Comparison of the intensity distributions of the deflection beam between experiment and simulation.

We repeat the measurements at 2.180 THz using QCL D7, but with its output beam filtered by a smaller 3 mm aperture. Figure 5(a) shows measured reflected beam using the mirror, while Figure 5(b) plots the beam diffracted by the metasurface. As expected, the diffracted beam is shifted with respect to the reflected one and is also focused. We then measure the 1D intensity distribution of the beams along the yellow dot lines in Figure 5(a) and (b), which are plotted together in Figure 5(c) for a comparison. The beam shifts from the position of -3.29 mm to 5.03 mm. The half power beam widths of reflected beam and diffracted beam are 3.04 mm and 0.45 mm, respectively. We also plot diffracted/focused beam in detail to allow for a comparison between measurement and simulation in Figure 5(d). We continue to characterize the metasurface at 2.188 THz, using the third laser QCL D8, applying an even smaller aperture of 2.5 mm to filter the beam, otherwise the beam is not circular enough. All the results, similar to the other two frequencies, are summarized in Figure 5(e)-(h). The beam shifts from -3.33 mm to 5.01 mm. The half power beam width of reflected beam and diffracted beam are 2.69 mm and 0.59 mm, respectively. We notice two anomalies here: One is the unexpected weak peak at the position of 2.9 mm in Figure 5(g), the reason for which is unclear, and other is the beam width in Figure 5(h), which is considerably larger than the modelling, the reason for which is the smaller input beam size.

We plot diffracted beams at the three frequencies together in Figure 5(i) to quantify the frequency dependence. We offset the peak position of the beam profile for 2.180 THz to zero in $x$-axis. Then, the relative peak position for 2.150 THz is 60 μm in experiment, while 45 μm predicted in simulation. The peak position of 2.188 THz is -20 μm in experiment, while -17 μm in simulation. With respect to the peak positions, experimental values agree reasonably well with the simulations. Furthermore, since we can observe the position difference when two lights around 2.188 THz with only a difference of 8 GHz, we seem to achieve a spectral resolution R of 273. However, suppose these three coherent signals are simultaneously shined onto the metasurface structure, it might be difficult to discriminate them as suggested by Figure 5(i). In this case we can introduce the spectrum inversion algorithm to resolve these lines. The details of the spectrum inversion algorithm are described in **Supporting Information Section 6**. To apply this method to our experimental data, we sum three intensity profile data by assuming the incident power to be distributed with a share of 0.2, 0.5, and 0.3 for the measured, but normalized beam profile at 2.150 THz, 2.180 THz, 2.188 THz in Figure 5 (i). It leads to summed intensity profile contributed by three different frequencies in Figure 5(j). We also assume a signal-to-noise (S/N) ratio of 30 dB for our measurements. By applying this technique, similar to performing frequency decomposition, we find the power share to be 0.209 for 2.150 THz. Compared to the original 0.2, the difference or error is 4.5%. The power share is 0.468 for 2.180 THz with an error of 6.4%, and 0.326 for 2.188 THz with an error 8.7%. Thus, the max uncertainty is 8.7% and is reasonably small. The comparison of the ideal and predicted intensity distribution for each frequency is also shown in Figure 5(j). Because we are able to differentiate two lines separated by 8 GHz at 2.19 THz, the realized resolution $R\ is\ at\ least$ 273.

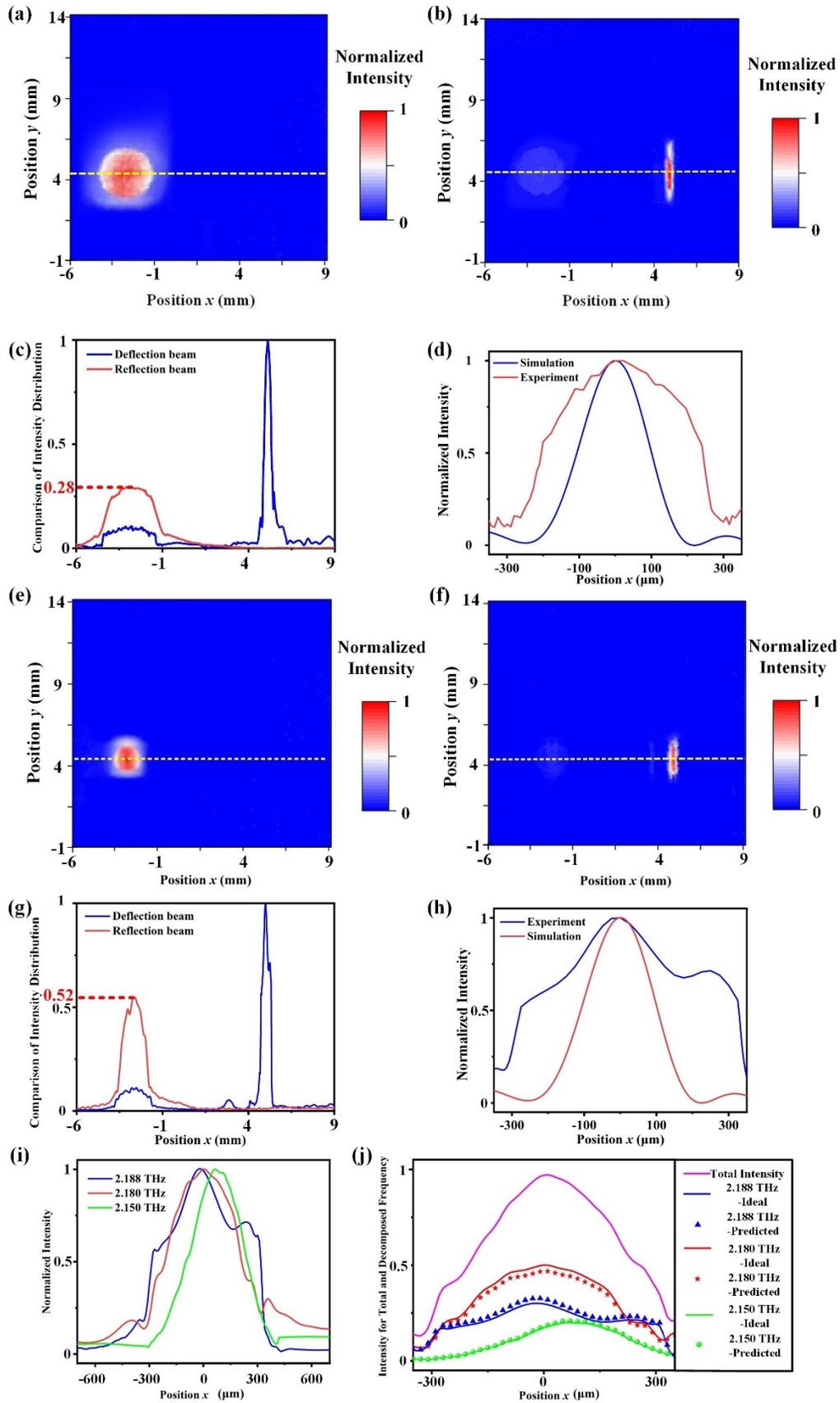

Figure 5. Measurement results of the metasurface spectrometer structure at 2.180 and 2.188 THz. (a) Reflection beam (b) deflection beam at 2.180 THz (c) Detailed intensity distributions of the reflection- and deflection-beam in 1D scanned along the dashed lines in (a) and (b). (d) Comparison of the intensity distributions of the deflection beam between experiment and simulation at 2.180 THz. (e) Reflection beam at 2.188 THz. (f) Deflection beam. (g) Detailed intensity distributions of the reflection- and deflection-beam in 1D scanned along the dashed lines in (e) and (f). (h) Comparison of the intensity distributions of the deflection beam between experiment and simulation at 2.188 THz. (i) Intensity distributions of the deflection beams for the three frequencies. (j) The total intensity distribution mixed with normalized intensity distributions for 2.150 THz, 2.180 THz, 2.188 THz from (i), but weighted with a factor of 0.2, 0.5, and 0.3, respectively. Also plotted are the three intensity distributions derived by the spectrum inversion method together with the normalized, weighted intensity distributions used in (j)

## 5, Conclusion and outlook

In summary, we propose and demonstrate a compact THz spectrometers based on a metasurface, which functions both in diffraction and focusing. We model, design and fabricate a metasurface to work in the frequency band between 1.7 and 2.5 THz. We have used three quantum cascaded lasers, emitting slightly different frequencies around 2.1 THz, to verify our concept, in particular, the spectral resolution. The diffracted beams can show different positions in the focusing plane even for two laser lines with the 8 GHz difference. This result implies a spectral resolution R of 273. However, to fully resolve them, we apply the spectral inversion method to confirm a R at least 273. The efficiency of the metasurface spectometer is found to be 84 % at 2.15 THz. The results suggest that we succeeded for the first time in demonstrating a compact spectrometer using metasurface at THz. Such a novel spectrometer has potential to replace classic, bulky grating spectrometers highly demanded for space observatories [4]-[6] and the same concept can be applied to other wavelengths. e.g. medium- IR (5-30 µm) grating spectrometer for the JWST-MIRI instrument [44]. It can also miniaturize THz spectrometers for many industrial and laboratory applications in the near future.

To further increase the spectral resolution, one needs to optimize the design of the metasurface to enhance the dispersion and to adjust NA properly. Another step is to explore a 2D metasurface to improve the intensity in the center of focus beam because the 2D metasurface has the effect of 2D focus, which increases the S/N ratio. Finally, the spectrometer system should be integrated with ultra-sensitive cryogenic detector arrays [46], [47] for space observatories, while with a suitable room temperature 2D detector array [45] for non-space applications,

## Acknowledgement

This work is supported by NWO-TTW Perspectief program (P15-36) "Free-form scattering optics", by EU RADIOBLOCKS project (grant number 101093934), and by TUDelft Space Institute. We thank Roland Horsten, Thomas Scholte, and Thim Zuidwijk for their helps to build the experimental set-up. We also thank Wenxiu Wang for supporting programming work and Yawen Wang in University of Electronic Science and Technology, Chengdu, China, for drawing 3D figures.

## Competing interests

Wilt Kao, and Alan Lee work in LongWave Photonics LLC, USA.

# Supporting Information

## Section 1: The variable parameters α and β for all the unit cells in the metasurface sample.

| No. | 1 | 2 | 3 | 4 | 5 | 6 | 7 | 8 | 9 | 10 | 11 | 12 | 13 | 14 | 15 | 16 | 17 | 18 | 19 | 20 |
|---|---|---|---|---|---|---|---|---|---|---|---|---|---|---|---|---|---|---|---|---|
| $\alpha$ | 116 | 72 | 123 | 78 | 129 | 83 | 135 | 87 | 140 | 90 | 25 | 92 | 147 | 94 | 33 | 96 | 36 | 97 | 37 | 97 |
| $\beta$ | 45 | -45 | -45 | 45 | 45 | -45 | -45 | 45 | 45 | -45 | 45 | 45 | 45 | -45 | 45 | 45 | -45 | -45 | 45 | 45 |
| No. | 21 | 22 | 23 | 24 | 25 | 26 | 27 | 28 | 29 | 30 | 31 | 32 | 33 | 34 | 35 | 36 | 37 | 38 | 39 | 40 |
| $\alpha$ | 37 | 97 | 36 | 96 | 34 | 95 | 148 | 93 | 26 | 90 | 21 | 87 | 137 | 84 | 131 | 79 | 124 | 73 | 117 | 66 |
| $\beta$ | -45 | -45 | 45 | 45 | -45 | -45 | -45 | 45 | -45 | -45 | 45 | 45 | 45 | -45 | -45 | 45 | 45 | -45 | -45 | 45 |
| No. | 41 | 42 | 43 | 44 | 45 | 46 | 47 | 48 | 49 | 50 | 51 | 52 | 53 | 54 | 55 | 56 | 57 | 58 | 59 | 60 |
| $\alpha$ | 110 | 56 | 103 | 42 | 96 | 27 | 89 | 135 | 80 | 123 | 69 | 111 | 54 | 100 | 34 | 90 | 135 | 79 | 120 | 64 |
| $\beta$ | 45 | -45 | -45 | 45 | 45 | -45 | -45 | -45 | 45 | 45 | -45 | -45 | 45 | 45 | -45 | -45 | -45 | 45 | 45 | -45 |
| No. | 61 | 62 | 63 | 64 | 65 | 66 | 67 | 68 | 69 | 70 | 71 | 72 | 73 | 74 | 75 | 76 | 77 | 78 | 79 | 80 |
| $\alpha$ | 105 | 41 | 92 | 137 | 79 | 118 | 60 | 101 | 32 | 87 | 128 | 70 | 107 | 42 | 91 | 132 | 73 | 109 | 43 | 91 |
| $\beta$ | -45 | 45 | 45 | 45 | -45 | -45 | 45 | 45 | -45 | -45 | -45 | 45 | 45 | -45 | -45 | -45 | 45 | 45 | -45 | -45 |
| No. | 81 | 82 | 83 | 84 | 85 | 86 | 87 | 88 | 89 | 90 | 91 | 92 | 93 | 94 | 95 | 96 | 97 | 98 | 99 | 100 |
| $\alpha$ | 131 | 70 | 106 | 36 | 87 | 124 | 62 | 99 | 21 | 78 | 112 | 44 | 89 | 125 | 62 | 98 | 138 | 73 | 106 | 30 |
| $\beta$ | -45 | 45 | 45 | -45 | -45 | -45 | 45 | 45 | -45 | -45 | -45 | 45 | 45 | 45 | -45 | -45 | -45 | 45 | 45 | -45 |
| No. | 101 | 102 | 103 | 104 | 105 | 106 | 107 | 108 | 109 | 110 | 111 | 112 | 113 | 114 | 115 | 116 | 117 | 118 | 119 | 120 |
| α | 81 | 113 | 41 | 86 | 119 | 50 | 90 | 123 | 55 | 92 | 126 | 57 | 93 | 126 | 57 | 92 | 125 | 54 | 90 | 121 |
| β | -45 | -45 | 45 | 45 | 45 | -45 | -45 | -45 | 45 | 45 | 45 | -45 | -45 | -45 | 45 | 45 | 45 | -45 | -45 | -45 |
| No. | 121 | 122 | 123 | 124 | 125 | 126 | 127 | 128 | 129 | 130 | 131 | 132 | 133 | 134 | 135 | 136 | 137 | 138 | 139 | 140 |
| α | 47 | 86 | 116 | 38 | 81 | 108 | 144 | 73 | 101 | 134 | 62 | 92 | 121 | 43 | 82 | 108 | 21 | 68 | 95 | 124 |
| β | 45 | 45 | 45 | -45 | -45 | -45 | -45 | 45 | 45 | 45 | -45 | -45 | -45 | 45 | 45 | 45 | -45 | -45 | -45 | -45 |
| No. | 141 | 142 | 143 | 144 | 145 | 146 | 147 | 148 | 149 | 150 | 151 | 152 | 153 | 154 | 155 | 156 | 157 | 158 | 159 | 160 |
| α | 45 | 82 | 106 | 138 | 63 | 91 | 116 | 30 | 72 | 96 | 123 | 40 | 77 | 100 | 127 | 45 | 79 | 100 | 127 | 44 |
| β | 45 | 45 | 45 | 45 | -45 | -45 | -45 | 45 | 45 | 45 | 45 | -45 | -45 | -45 | -45 | 45 | 45 | 45 | 45 | -45 |
| No. | 161 | 162 | 163 | 164 | 165 | 166 | 167 | 168 | 169 | 170 | 171 | 172 | 173 | 174 | 175 | 176 | 177 | 178 | 179 | 180 |
| α | 78 | 99 | 124 | 38 | 73 | 95 | 117 | 26 | 64 | 88 | 108 | 133 | 49 | 78 | 97 | 119 | 26 | 62 | 85 | 103 |
| β | -45 | -45 | -45 | 45 | 45 | 45 | 45 | -45 | -45 | -45 | -45 | -45 | 45 | 45 | 45 | 45 | -45 | -45 | -45 | -45 |
| No. | 181 | 182 | 183 | 184 | 185 | 186 | 187 | 188 | 189 | 190 | 191 | 192 | 193 | 194 | 195 | 196 | 197 | 198 | 199 | 200 |
| α | 125 | 34 | 67 | 87 | 104 | 126 | 34 | 66 | 86 | 102 | 122 | 26 | 58 | 80 | 96 | 113 | 134 | 42 | 69 | 86 |
| β | -45 | 45 | 45 | 45 | 45 | 45 | -45 | -45 | -45 | -45 | -45 | 45 | 45 | 45 | 45 | 45 | 45 | -45 | -45 | -45 |
| No. | 201 | 202 | 203 | 204 | 205 | 206 | 207 | 208 | 209 | 210 | 211 | 212 | 213 | 214 | 215 | 216 | 217 | 218 | 219 | 220 |
| α | 101 | 118 | 139 | 47 | 71 | 87 | 100 | 117 | 136 | 40 | 65 | 82 | 95 | 108 | 125 | 23 | 48 | 69 | 84 | 96 |
| β | -45 | -45 | -45 | 45 | 45 | 45 | 45 | 45 | 45 | -45 | -45 | -45 | -45 | -45 | -45 | 45 | 45 | 45 | 45 | 45 |
| No. | 221 | 222 | 223 | 224 | 225 | 226 | 227 | 228 | 229 | 230 | 231 | 232 | 233 | 234 | 235 | 236 | 237 | 238 | 239 | 240 |
| α | 108 | 123 | 140 | 42 | 63 | 78 | 90 | 100 | 112 | 127 | 21 | 42 | 62 | 76 | 87 | 96 | 106 | 118 | 130 | 24 |
| β | 45 | 45 | 45 | -45 | -45 | -45 | -45 | -45 | -45 | -45 | 45 | 45 | 45 | 45 | 45 | 45 | 45 | 45 | 45 | -45 |
| No. | 241 | 242 | 243 | 244 | 245 | 246 | 247 | 248 | 249 | 250 | | | | | | | | | | |
| α | 43 | 60 | 73 | 83 | 92 | 100 | 109 | 119 | 130 | 21 | | | | | | | | | | |
| β | -45 | -45 | -45 | -45 | -45 | -45 | -45 | -45 | -45 | 45 | | | | | | | | | | |

## Section 2: Co-polarization reflection wave property of a unit cell and the intensity distribution in the focusing plane.

For a co-polarization reflection beam, we give an example of a unit cell working at 2.1 THz in Figure S1(a). We can show for different parameters for the unit cells, the phase curve does not have the same property as the cross-polarization. The reflection amplitude is always low and is less than 0.2 across the whole frequency span, while it is more than 0.8 for the cross-polarization (see Figure 2(e)). In Figure S1(b), we give several parameter curves of the phase, $\text{Arg}(r_{yy})$, and amplitude, $|r_{yy}|$, as a function of frequency for unit cells with parameters $\beta$ (-45° and 45°) and $\alpha$ (20° and 46°). From 1.7 to 2.5 THz, for all the phase curves, they are not parallel to each other anymore, so they don't have the fixed phase relationship as the cross-polarization reflection wave in Figure 2(f). In addition, in the whole band, the amplitude from every reflection curve is always lower than 0.2. At last, we give a comparison of intensity distributions between cross-polarization and co-polarization deflected light in the focusing plane. In Figure S1(d), we can see the co-polarization does not have a focused, deflected beam, but spread its intensity everywhere in the focusing plane. In addition, the intensity is much lower compared with the cross-polarization diffracted beam, especially in the focus area.

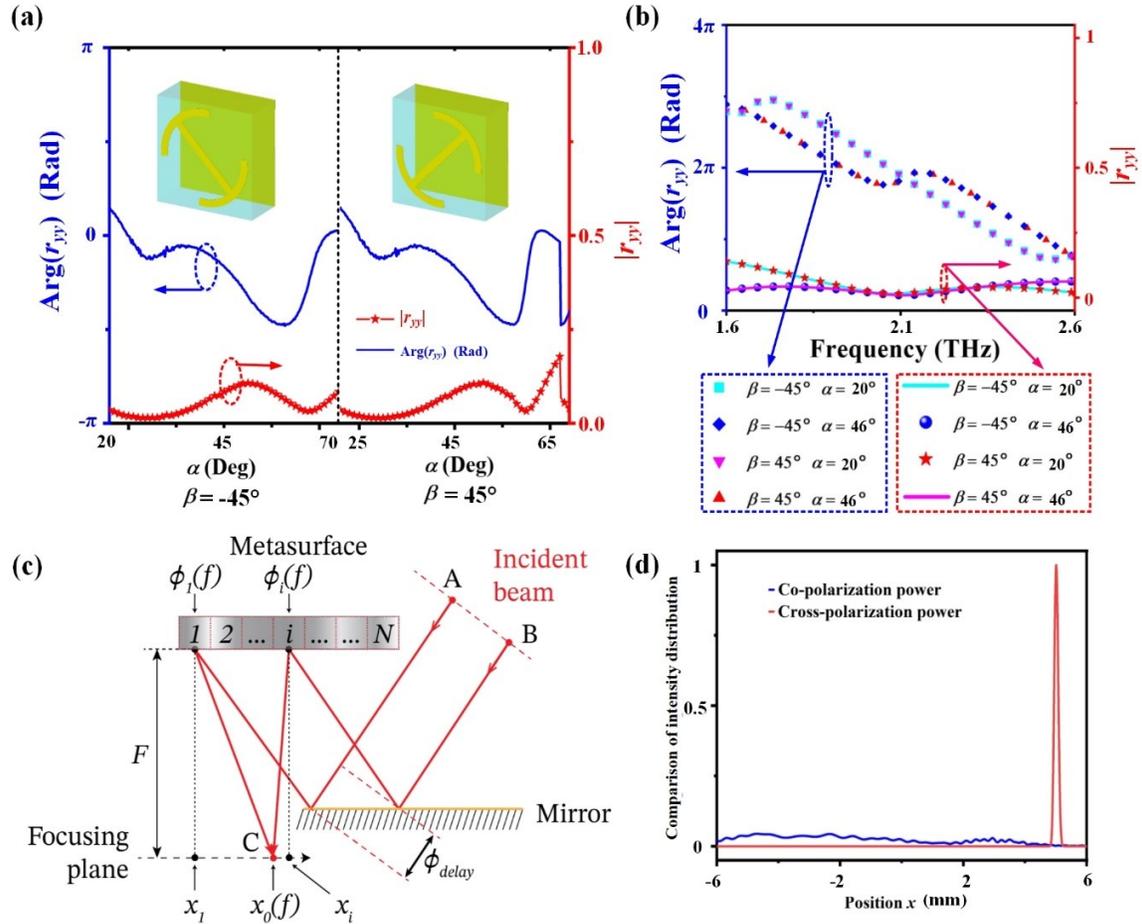

Figure S1 (a) Properties of the phase, Arg($r_{yy}$), and amplitude, |$r_{yy}$|, associating with a co-polarization deflection beam, for two fixed $\beta$ values (-45° and 45°), but with a variable $\alpha$ for different unit cells at 2.1 THz. (b) Phase, Arg($r_{yy}$), and amplitude, |$r_{yy}$|, as a function of frequency for unit cells with parameters $\beta$ (-45° and 45°) and $\alpha$ (20° and 46°). (c) Schematic of final design for metasurface spectrometer system. (d) Comparison of intensity distributions between cross-polarization and co-polarization deflected light in the focusing plane.

## Section 3: QCLs

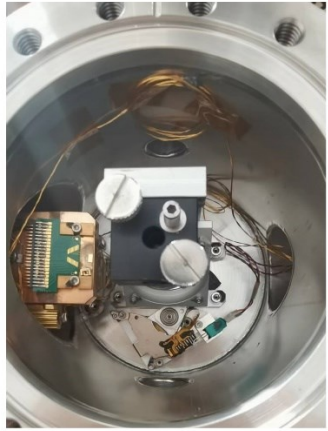

(a)

Quantum Cascaded
Laser inside Cooler

(b)

| Laser Number | D4 | D7 | D8 |
|---|---|---|---|
| Frequency (THz) | 2.150 | 2.180 | 2.188 |
| Current (mA) | 179 | 175 | 173 |
| Working Voltage (V) | 14.4 | 14.4 | 14.5 |
| Average Power (mW) | 0.78 | 0.64 | 0.72 |

Figure S2, (a) Photo of the QCL chip mounted inside a CryoTel Stirling cryocooler from Sunpower Inc (b) Detailed information of 3 QCLs used for our experiment. They were operated at a temperature of 49K. Since the window of the cooler is made of ultra-high molecular-weight polyethylene (UHMW-PE), which has a transmission roughly 90 %, the effective power of a QCL is 10 % lower than what indicated in the table [1].

## Section 4: Shaping the QCL beams

In Figure S3(b) shows interior structure of optical components inside the cooler. The emission from a QCL (green part) first passes through some optical elements, such as a parabolical reflector and beam splitter (red part). If the phase centre of the beam is not in focus spot of the parabolical reflector, the reflected beam will be distorted and no longer a Gaussian beam if the original beam is ideal. We have to notify that the optics inside the cooler was originally designed for a different QCL operated at 4.7 THz [1]. Figure S3(a) shows a nonideal beam of QCL D4 operated at 2.150 THz, measured outside the cooler. In order to reach a circular beam which is crucial for our experiment, we introduce a diaphragm with an aperture of 4 mm in diameter. The diaphragm is 1 cm away from the window of the cooler. Then, the imaging plane to map the beam pattern is 1 cm after the diaphragm. The setup is shown in Figure S3(d). The filtered beam is shown in Figure S3(c), which is much improved than the previous one. We further optimize it by changing the aperture size and adjusting the optical elements inside the cooler. The final optimized beam is described in the manuscript. Furthermore, the original beams for QCL D7 2.180 THz and D8 2.188 THz are shown in Figure S3(e) and (f), respectively.

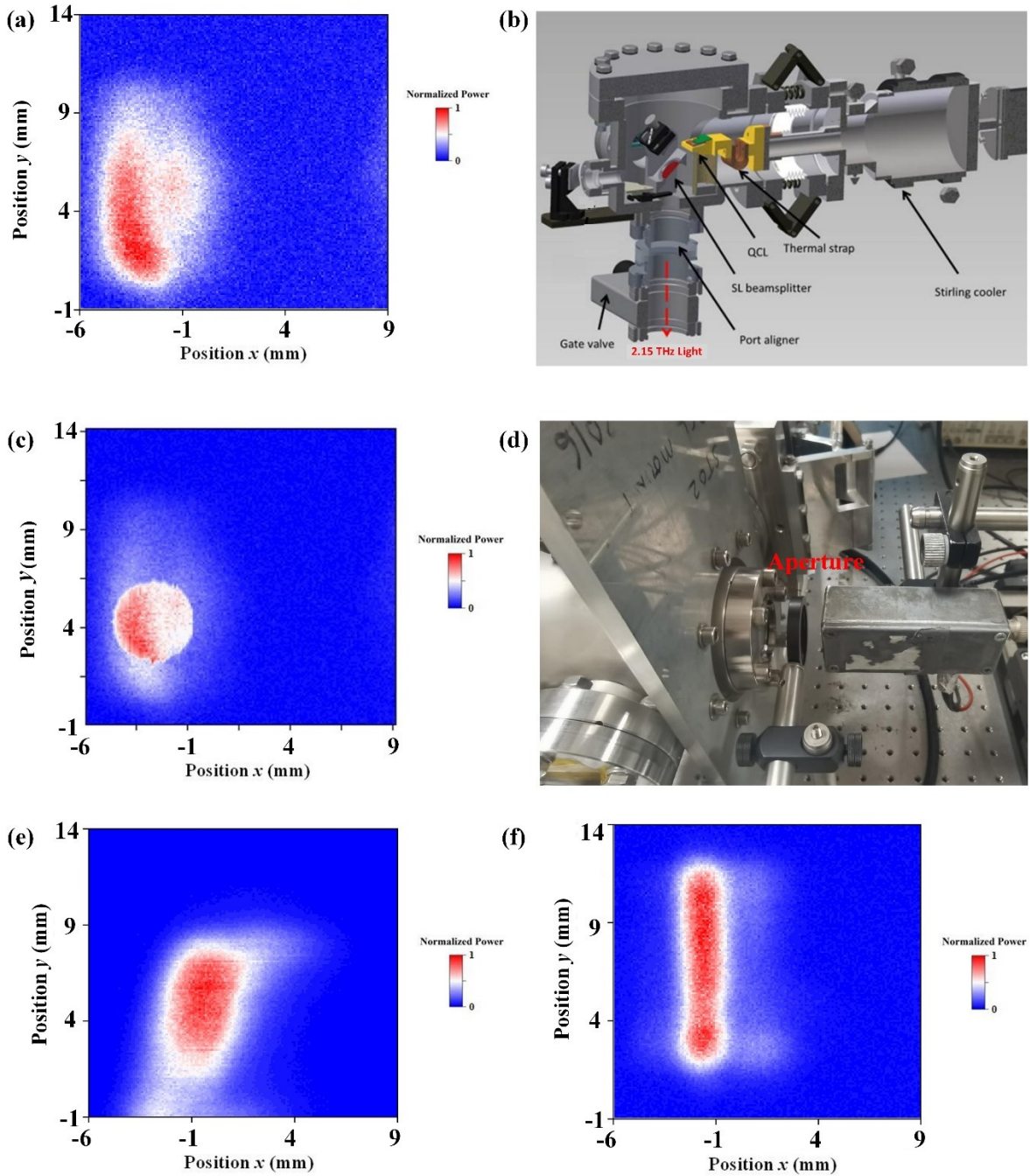

Figure S3 (a) Original beam pattern of QCL D4, operated at 2.150 THz. (b) Interior structure of the cooler [1]. (c) Filtered beam after the diaphragm. (d) The setup of adding the diaphragm. Original beam patterns of QCL D7 at 2.180 THz in (e) and D8 2.188 THz in (f).

## Section 5: Intensity distribution in the focusing plane for different incident beam size at 2.10 THz.

Figure S4 (b) shows the intensity distributions of a diffraction beam in the focusing plane in Figure S4 (a) when the diameter of the incident beam of the 2.10 THz light varies from 11.25 mm to 2 mm. We ensure that the light is incident on the central area of the metasurface. From the results in (b), we find that the effects of focusing and deflection are not affected when the size of incident beam is varied, and the peaks of different diffraction beams are almost at the same position. The only change is that beam size, which increases as the incident beam size decreases. When the size of incident beam is 11.25mm, 4mm, 3mm, 2.5mm, and 2mm, the corresponding half power beam width of the diffraction beam is 162 μm, 357μm, 439μm, 575μm, and 764μm, respectively.

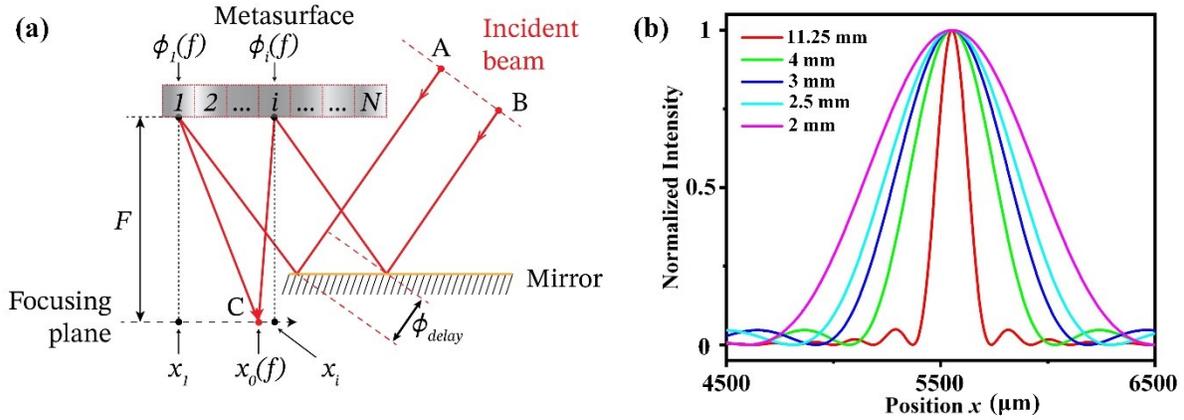

Figure S4 (a) Schematic of the focusing plane by the dash line in the metasurface spectrometer configuration. (b) Intensity distribution in the focusing plane for different incident beam size for the light at 2.10 THz.

## Section 6: Spectrum Inversion Method

If we know the intensity distribution of a diffracted beam in the focusing plane, we can determine the wavelength and the intensity of the light. First, we calibrate the metasurface spectrometer by measuring the intensity distribution $U(\lambda_i, x_j)$ in the focusing plane for incident wavelength $\lambda_i$ with unit intensity 1. $x_j$ is a position in the focusing plane along $x$ direction in Figure S5(a). If we give an incident wavelength $\lambda_i$ with intensity $I_i$, the intensity in position $x_j$ can be expressed as $V(\lambda_i, x_j) = I_i \cdot U(\lambda_i, x_j)$, which is shown in Figure S5(c). If the incident light is multi-wavelengths ($\lambda_i$, $i$ is integer from 1 to $N$), the total intensity in $x_j$ is, $V_{SUM}(x_j) = I_1 U(\lambda_1, x_j) + I_2 U(\lambda_2, x_j) + I_3 U(\lambda_3, x_j) + \cdots I_3 U(\lambda_N, x_j) = \sum_{i=1}^{N} I_i U(\lambda_i, x_j)$. We can measure the intensity in different position ($x_j$, $j$ is integer from 1 to $M$) in focusing plane. Now, the problem becomes to solve a system of linear equations:

$$V_{SUM}(x_j) = I_1 U(\lambda_1, x_j) + I_2 U(\lambda_2, x_j) + I_3 U(\lambda_3, x_j) + \cdots I_3 U(\lambda_N, x_j) \tag{1}$$

where $j$ is integer from 1 to $M$. $U(\lambda_i, x_j)$ and $V_{SUM}(x_j)$ are known, which we can obtain by measurement. $I_i$ ($i$ is integer from 1 to $N$) is what we want by solving the linear equations as long as $M \geq N$. We write the system in matrix form as follows,

$$\begin{bmatrix} V(x_1) \\ V(x_2) \\ V(x_3) \\ \cdots \\ V(x_M) \end{bmatrix} = \begin{bmatrix} U(\lambda_1, x_1) & U(\lambda_2, x_1) & U(\lambda_3, x_1) & \cdots & U(\lambda_N, x_1) \\ U(\lambda_1, x_2) & U(\lambda_2, x_2) & U(\lambda_3, x_2) & \cdots & U(\lambda_N, x_2) \\ U(\lambda_1, x_3) & U(\lambda_2, x_3) & U(\lambda_3, x_3) & \cdots & U(\lambda_N, x_n) \\ \cdots & \cdots & \cdots & \cdots & \cdots \\ U(\lambda_1, x_M) & U(\lambda_2, x_M) & U(\lambda_3, x_M) & \cdots & U(\lambda_N, x_M) \end{bmatrix} \begin{bmatrix} I_1 \\ I_2 \\ I_3 \\ \cdots \\ I_N \end{bmatrix}, \tag{2}$$

which can be simplified as $V = U \cdot I$. In a realistic measurement system, there is noise in $V$. Thus, the problem is to find $I$ to ensure $\min_I \{\| UI - V \|_2^2\}$. We adopt Tikhonov regularization to solve this minimization problem [2]. The general form of Tikhonov regularization is expressed as,

$$I_\Omega = \min_I \{\| UI - V \|_2^2 + \Omega^2 \| I \|_2^2\} \tag{3}$$

where $\Omega$ is a regularization parameter. We note that if $\Omega$ is too large, it will then not fit the measurement data $V$ properly and the residual will be too large. In contrast, if $\Omega$ is too small, the solution will then be dominated by the contributions from the noise in $V$ and the solution norm will be large. As a result, we should find the proper regularization parameter $\Omega$ to get the good solution.

We use L-curve method to choose a proper $\Omega$ [3]. First, we use singular value decomposition to decompose the matrix $U$ ($M \times N, M \geq N$) in the form,

$$U = \sum_{i=1}^{N} q_i \sigma_i s_i^T \qquad (4)$$

where the left and right singular vectors $q_i$ and $s_i$ are orthonormal, and the singular values $\sigma_i$ are nonnegative values that are in non-increasing order, namely, $\sigma_1 \geq \sigma_2 \geq \cdots \geq \sigma_N \geq 0$. Then, the Tikhonov solution is,

$$I_\Omega = \sum_{i=1}^{N} t_i \frac{q_i^T V}{\sigma_i} s_i \qquad (5)$$

where $t_1, \cdots, t_N$ are the Tikhonov filter factors, which depend on $\sigma_i$ and $\Omega$ as,

$$t_i = \frac{\sigma_i^2}{\sigma_i^2 + \Omega^2} \approx \begin{cases} 1 & \sigma_i^2 \gg \Omega \\ \sigma_i^2/\Omega^2, & \sigma_i^2 \ll \Omega \end{cases} \qquad (6)$$

Then, the solution and residual norms are,

$$\| I_\Omega \|_2^2 = \sum_{i=1}^{N} (t_i \frac{q_i^T V}{\sigma_i})^2 \qquad (7)$$

$$\| U I_\Omega - V \|_2^2 = \sum_{i=1}^{N} ((1-t_i) q_i^T V)^2 \qquad (8)$$

L-curve is plotted with horizontal axis $\| U I_\Omega - V \|_2^2$ and vertical axis $\| I_\Omega \|_2^2$.

With the spectrum inversion method we discussed previously, we plot the L-curve in Figure S5(b) and select the regularization parameter $\| I_\Omega \|_2^2$ 0.0612 at the corner of the curve. To apply this method to our data in our manuscript, we sum three intensity profile data by assuming the incident power to be distributed with a share of 0.2, 0.5, and 0.3 for the measured, but normalized beam profile at 2.150 THz, 2.180 THz, 2.188 THz in Figure 5 (i), that leads to summed intensity profile contributed by three different frequencies in Figure 5(j). We also need to assume a signal to noise ratio of 30 dB for our measurements. With the spectrum inversion method, we plot the L-curve in Figure S5(b) and select the regularization parameter $\| I_\Omega \|_2^2$ 0.0612 at the corner of the curve. By applying this technique, we find the power fraction to be 0.209 for 2.150 THz. Compared to the original 0.2, the difference or error is 4.5%. The power fraction is 0.468 for 2.180 THz, the error being 6.4%, and 0.326 for 2.188 THz with an error 8.7%. Thus, the max uncertainty is 8.7% and is still acceptably small. The comparison of ideal and predicted intensity distribution for each frequency is also shown in Figure 5(j). Since we are able to distinguish the 8 GHz difference at 2.19 THz, we have realized a resolution R of 273.

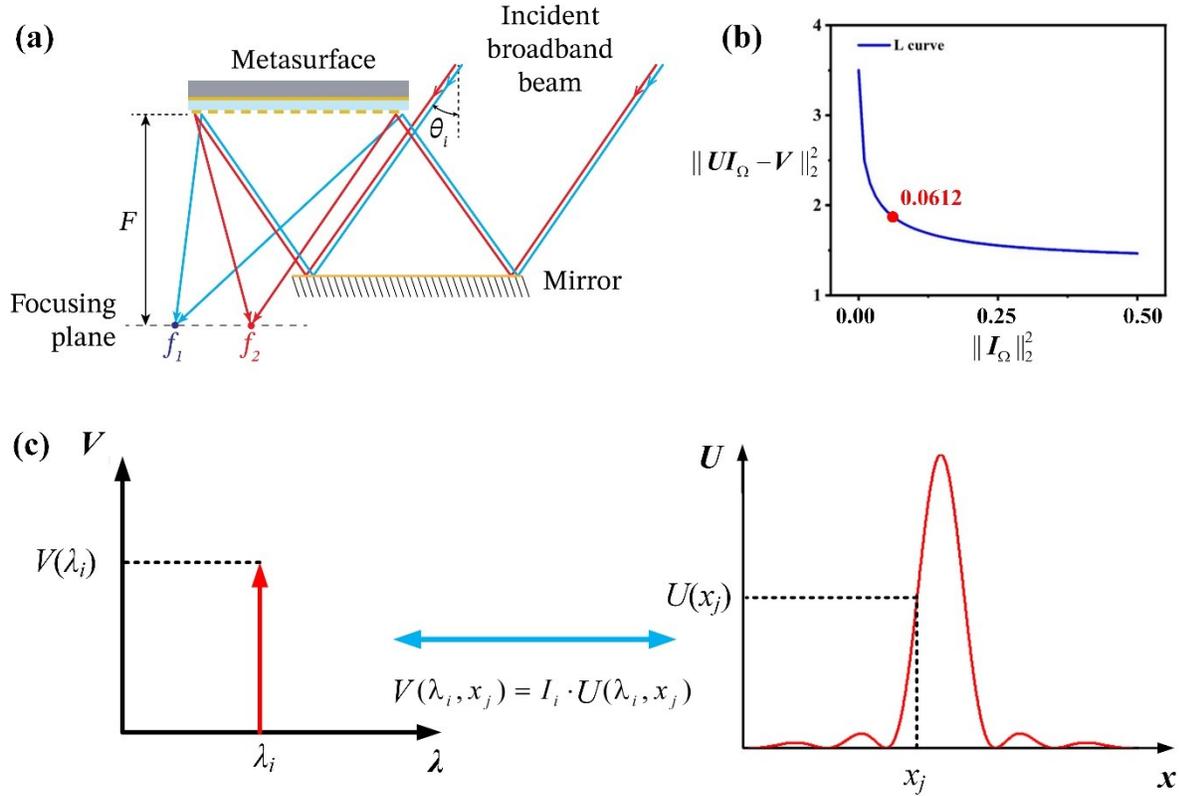

Figure S5. (a) Schematic of a metasurface spectrometer concept. (b) L-curve of our example. (c) The principle of spectrum inversion method for a single wavelength.